%% file: window_compare.tex
\documentclass[10pt, conference, compsocconf]{IEEEtran}
\usepackage{verbatim}
\pdfoutput=1

\usepackage[pdftex]{graphicx}
\ifCLASSINFOpdf
   \usepackage[pdftex]{graphicx}
\else
\fi
\begin{document}
%
\title{Bias voltage Control of Avalanche Photo-Diode Using a Window Comparator}


\author{\IEEEauthorblockN{Subash Sachidananda and Alexander Ling}
\IEEEauthorblockA{Centre for Quantum Technologies\\
National University of Singapore\\
Singapore - 117543\\
subash.jies@gmail.com, cqtalej@nus.edu.sg}
}


%


\maketitle

\begin{abstract}
This work aims at controlling the bias voltage of APDs, used for single photon detection, with a micro-controller through pulse height comparison.
\end{abstract}


%
\IEEEpeerreviewmaketitle

\input{Sections}
\bibliographystyle{IEEEtran}
%

\bibliography{win_compare}




\end{document}

%% file: Sections.tex
\section{Introduction}
Single photon detection is an essential tool for studying photon entanglement in several areas of research including Quantum Optics and Quantum Cryptography. The most popular method to detect single photons is by using Avalanche Photo Diodes (APD). The implementation of such detectors is done by applying a large reverse bias voltage ($V_{BIAS}$) to the APD, which is equal to or above its breakdown voltage ($V_{BR}$). When a single photon strikes the APD operating under the biased condition, a small current flows through the diode for a few nanoseconds and this can be picked up across a suitable load resistor to obtain a pulse with magnitude of a few hundred millivolts and width of a few nanoseconds.\\
To pursue long range entanglement studies, photon entanglement sources may need to be operated in novel environments, e.g. in space. To achieve this goal, we are working on a photon entanglement source designed to work in a pico-satellite in lower earth orbit ($\approx$500 km). However capture and processing of such tiny, high-speed pulses from the APD is quite tricky, considering the environmental and resource constraints encountered in pico-satellites and the limitations present in commercially available electronic components. 
\subsection{Motivation}
There exist several hard boundaries when it comes to deploying systems in lower earth orbit. The temperature is not stable and varies over a wide range every 100 minutes or so, with each revolution of the satellite around the earth. The energy available to run the components is limited in space and hence the power consumption of the hardware should be strictly within its allotted quota ($\approx$1.5 W). The volume of the satellite itself is limited and hence there are restrictions on the size and weight of the circuit board and the optics equipment.
The most notable impact of temperature variation in space is on the electronic components and especially on the breakdown voltage ($V_{BR}$) of the APDs, used for single photon detection. It is well known that $V_{BR}$ of an APD reduces with temperature. Hence the height of the output voltage pulse of the APD also varies with temperature. This means that the reverse bias voltage ($V_{BIAS}$) also has to be adjusted to ensure that $V_{BIAS} - V_{BR} = Constant$, at all times, in order to maintain the output voltage peak height between 500mV and 800mV. The output of the APDs can be stabilized by cooling them down to a known temperature and operating them at this fixed value so that $V_{BR}$ and $V_{BIAS}$ are held constant too. This is typically done with a thermoelectric cooling element, but is ruled out in our system from power considerations alone. Hence we propose to use a window comparator which can constantly monitor the pulse height of the output voltage of the APD. Temperature compensated $V_{BIAS}$ control circuits have been proposed in the past \cite{maeda} , but we use the Digital to Analogue Converter (DAC) on a micro-controller to dynamically control $V_{BIAS}$ of the APD based on the output of the window comparator. 
\section{Hardware set-up}
\begin{figure}[!t]
\centering
\includegraphics[width=3.0in, trim=0.5in 1in 0 0.5in]{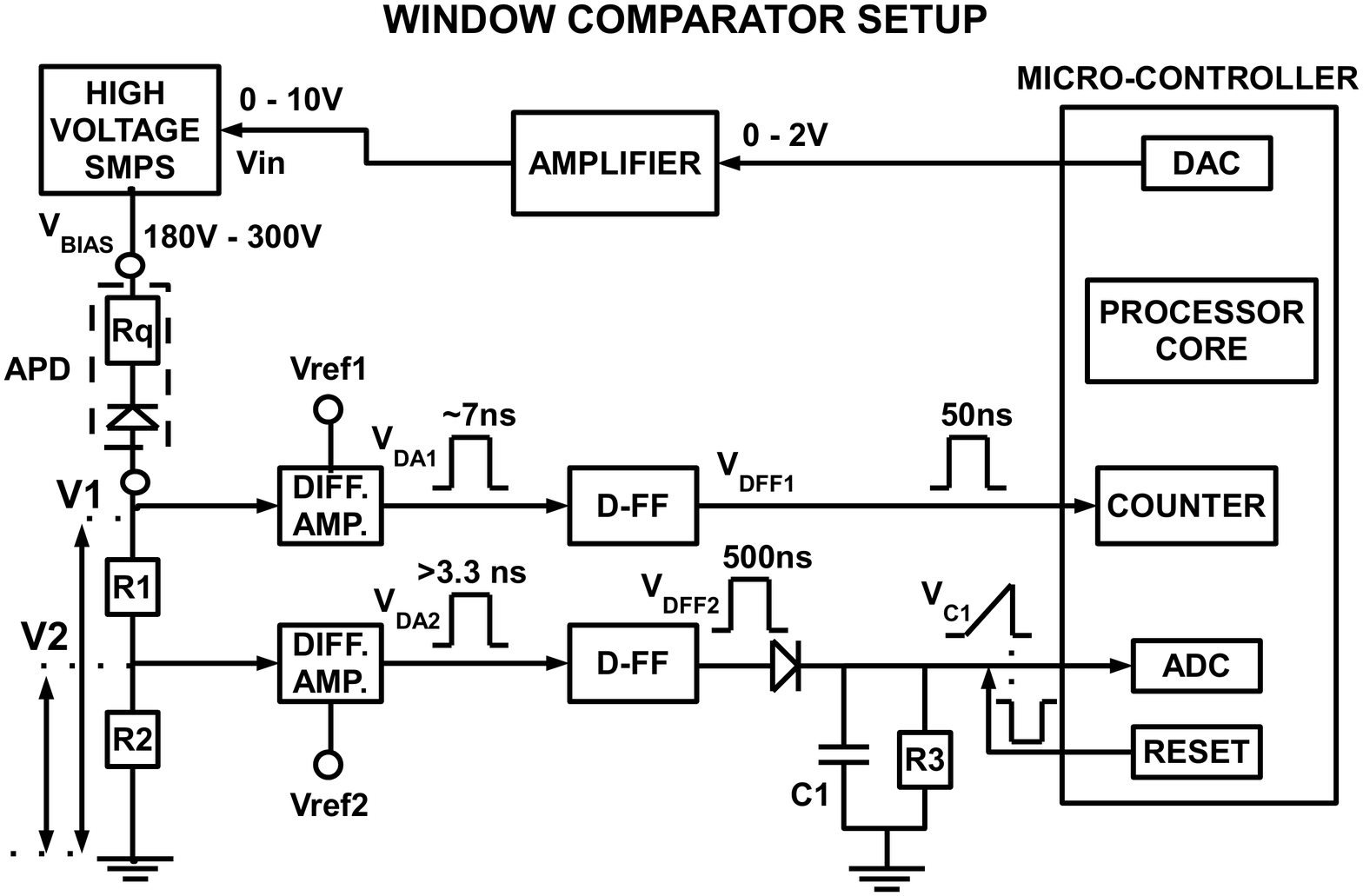}
\caption{Implementation of a window comparator in hardware}
\label{fig_WinCompare}
\end{figure}

The window comparator we propose to build is composed of several modules as shown in figure \ref{fig_WinCompare}.
\subsubsection*{Photon detection}
 We use the C30902 Perkin Elmer APD in passive quenching mode \cite{zappa}. The Matsusada-TS0.3P high voltage switch mode power supply (SMPS) is used to maintain the $V_{BIAS}$ for the APD between 180V and 300V DC which is incidentally the $V_{BR}$ range for the APD. The sense is split up into two resistors R1 and R2. Upon photon detection a small pulse of current sweeps past the sense which generate voltages V1 and V2 across (R1+R2) and R2 respectively.
\subsubsection*{Window comparator}
\begin{figure}[!t]
\centering
\includegraphics[width=3.0in, trim=0.5in 1in 0 0.5in]{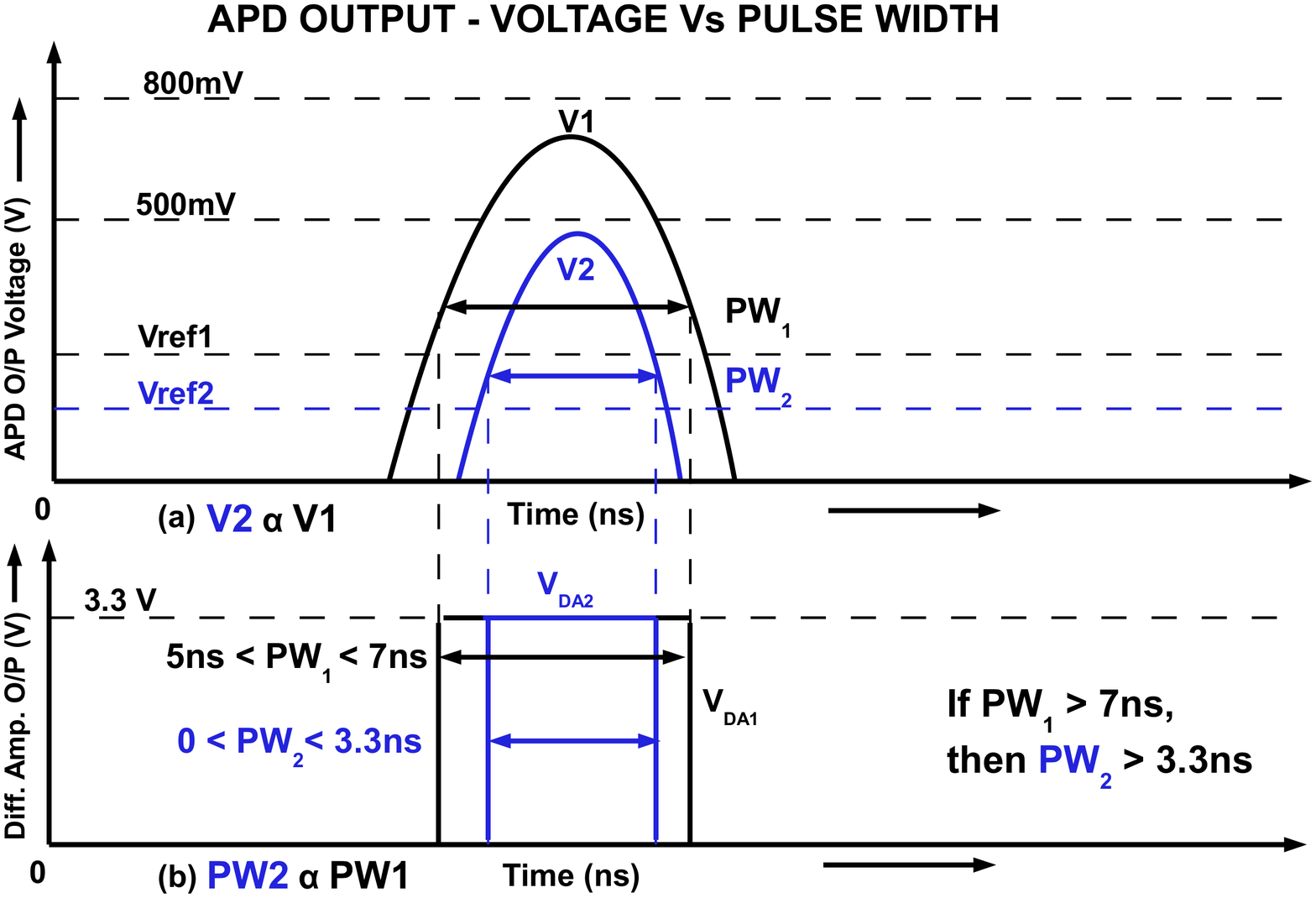}
\caption{Variation of pulse width with voltage (color online)}
\label{fig_VotlageVsWidth}
\end{figure}

The voltage pulses V1 and V2 are shown in figure \ref{fig_VotlageVsWidth}(a). Since R1 and R2 form a resistor ladder, V2 $\alpha$ V1 at all times. V1 and V2 are fed into a differential amplifier (DA) as shown in figure \ref{fig_WinCompare}, to convert them into square pulses of CMOS logic levels(0-3.3V). The DA compares V1 (V2) against fixed reference voltages Vref1 (Vref2) respectively and generates a high when V1(V2) is greater than Vref1(Vref2) by 100mV. The output of the differential amplifiers for V1 ($V_{DA1}$) and V2 ($V_{DA2}$) are depicted in figure \ref{fig_VotlageVsWidth}(b). Thus the pulse width of $V_{DA1}$ and $V_{DA2}$ depend directly on the pulse height of V1 and V2. For our system we seek to maintain the voltage V1 between 500mV and 800mV and we set Vref1 such that width of $V_{DA1}$ is between 5ns and 7ns for the voltage range of V1. Correspondingly V2 is also restricted and we set Vref2 such that the width of $V_{DA2}$ is between 0 and 3.3ns for the set range of V1. We intend to keep $V_{DA1}$ below 7ns to prevent accidental coincidences. When temperature changes cause $V_{BR}$ for the APD to decrease, the pulse height of V1 increases above 800mV and the width of $V_{DA2}$ is more than 3.3ns. 
\subsubsection*{Pulse stretching and shaping}
The output of differential amplifiers, $V_{DA1}$ and $V_{DA2}$, are passed through a D Flip-flop (D-FF) to stretch their widths to 50ns ($V_{DFF1}$) and 500ns ($V_{DFF2}$) respectively. The pulse stretching is important to ensure that slower hardware like micro-controller is able to process these high speed signals. We keep $V_{DFF2}$ fixed at width of 500ns and a height of 3.3V so that each pulse contains the same amount of energy. The flip-flops require that the input pulses be at least 3.3ns wide for proper functioning and thus $V_{DA1}$/$V_{DA2}$ need to be atleast 3.3ns to pass through the flip-flops. In this system, $V_{DA2}$ pulses less than 3.3ns wide never make it past the D-FF. This ensures that when APD pulses are greater than 800mv (overly high $V_{BIAS}$), the width of $V_{DA2}$ pulses increase above 3.3ns and are able to go past the D-FF.
\subsubsection*{Counting and voltage integration}
$V_{DFF1}$ is fed into a hardware counter present on the micro-controller chip. We use the PSoC3 controller chip from Cypress Semiconductors which can independently count up to $2^{32}$ pulses. $V_{DFF2}$ is fed into an integrator circuit consisting of a capacitor which charges up quickly (through the diode) and discharges slowly (through R3). As more pulses of $V_{DFF2}$ come in, the average voltage $V_{C1}$, increases across the capacitor. Since each pulse of $V_{DFF2}$ has the same energy, $V_{C1}$ is nearly proportional to the number of pulses. The ADC module of the PSoC3 is interfaced to read the capacitor voltage $V_{C1}$ and calculate the number of pulses. The PSoC3 can also reset the voltage $V_{C1}$ at any time by quickly discharging the capacitor thus making $V_{C1}=0V$. After a reset, the PSoC3 waits for a fixed time before sensing $V_{C1}$ with the ADC. This is useful to count the number of pulses that come in only within this fixed time frame (after reset and before ADC sense).
\subsubsection*{Feedback loop for $V_{BIAS}$ control} 
The output voltage $V_{BIAS}$ of the Matsusada SMPS can be controlled by changing voltage at the input pin (Vin) between 0 and 10V. The DAC of the PSoC3 provides dynamically varying voltages between 0 - 2V which can be set in software. The ouput of the DAC is thus amplified and fed into the Vin pin of the SMPS. Depending on $V_{DFF1}$ and $V_{DFF2}$, the software on the PSoC3 is able to estimate if $V_{BIAS}$ too high or too low. $V_{BIAS}$ is too high if $V_{C1}$ is consistently high (pulse rate of $V_{DFF2}$ is high) for a given count rate of $V_{DFF1}$ and the PSoC reduces $V_{BIAS}$ in steps until $V_{C1}$ is below a threshold. $V_{BIAS}$ is too low if count rate of $V_{DFF1}$ is very low and the PSoC increases $V_{BIAS}$ in steps until the count rate reaches above a threshold.
\section{Conclusion and Future work}
In this paper we proposed a micro-controller based feedback technique to adjust the bias voltage of an APD whose breakdown voltage varies frequently due to rapid temperature changes. The system we proposed is light weight and energy efficient and is suitable for deployment in compact optics experiments. It is quite flexible and can be calibrated to work at different voltage levels and pulse widths. We have designed, built and verified the proper working of all the individual modules described above. As part of future work, we intend to integrate all the hardware components on a single circuit board and demonstrate a complete and working system.

%% file: window_compare.bbl
\begin{thebibliography}{1}
\providecommand{\url}[1]{#1}
\csname url@samestyle\endcsname
\providecommand{\newblock}{\relax}
\providecommand{\bibinfo}[2]{#2}
\providecommand{\BIBentrySTDinterwordspacing}{\spaceskip=0pt\relax}
\providecommand{\BIBentryALTinterwordstretchfactor}{4}
\providecommand{\BIBentryALTinterwordspacing}{\spaceskip=\fontdimen2\font plus
\BIBentryALTinterwordstretchfactor\fontdimen3\font minus
  \fontdimen4\font\relax}
\providecommand{\BIBforeignlanguage}[2]{{%
\expandafter\ifx\csname l@#1\endcsname\relax
\typeout{** WARNING: IEEEtran.bst: No hyphenation pattern has been}%
\typeout{** loaded for the language `#1'. Using the pattern for}%
\typeout{** the default language instead.}%
\else
\language=\csname l@#1\endcsname
\fi
#2}}
\providecommand{\BIBdecl}{\relax}
\BIBdecl

\bibitem{maeda}
\BIBentryALTinterwordspacing
J.~S. M. N.~J. Maeda, Masaaki~(Tokyo, ``Bias voltage control circuitry for
  avalanche photodiode taking account of temperature slope of breakdown voltage
  of the diode, and method of adjusting the same,'' Patent 6\,157\,022,
  December, 2000. [Online]. Available:
  \url{http://www.freepatentsonline.com/6157022.html}
\BIBentrySTDinterwordspacing

\bibitem{zappa}
S.~{Cova}, M.~{Ghioni}, A.~{Lacaita}, C.~{Samori}, and F.~{Zappa}, ``Avalanche
  photodiodes and quenching circuits for single-photon detection,'' vol.~35,
  pp. 1956--+, apr 1996.

\end{thebibliography}
